\documentclass{article}
\usepackage[T1]{fontenc} % add special characters (e.g., umlaute)
\usepackage[utf8]{inputenc} % set utf-8 as default input encoding
\usepackage{ismir,amsmath,cite,url}
\usepackage{tabularx}
\usepackage{booktabs}
\usepackage{graphicx}
\usepackage{multicol}
\usepackage{multirow}
\usepackage[bookmarks=false,hidelinks]{hyperref}
\usepackage{caption}
\usepackage{subcaption}
\usepackage{color}
\usepackage{enumitem}

\usepackage{lineno}
\title{Improving Choral Music Separation through Expressive Synthesized Data from Sampled Instruments}

\multauthor
{Ke Chen$^1$ \hspace{.3cm} Hao-Wen Dong$^1$ \hspace{.3cm} Yi Luo$^2$ \hspace{.3cm} Julian McAuley$^1$ \vspace{.2cm}} {\textbf{Taylor Berg-Kirkpatrick}$^1$  \hspace{.3cm} \textbf{Miller Puckette}$^1$ \hspace{.3cm} \textbf{Shlomo Dubnov}$^1$ \vspace{.2cm} \\
$^1$UC San Diego, USA \hspace{.3cm} $^2$Tencent AI Lab, China \\
{\tt\small$^1$\{knutchen, hwdong, jmcauley, tberg, msp, sdubnov\}@ucsd.edu  $^2$oulyluo@tencent.com} \vspace{-0.3cm}
}

\def\authorname{K. Chen et al.}

\usepackage{amssymb}

\begin{document}

\maketitle
\begin{abstract}
Choral music separation refers to the task of extracting tracks of voice parts (e.g., soprano, alto, tenor, and bass) from mixed audio. The lack of datasets has impeded research on this topic as previous work has only been able to train and evaluate models on a few minutes of choral music data due to copyright issues and dataset collection difficulties. In this paper, we investigate the use of synthesized training data for the source separation task on real choral music. 
We make three contributions:
first, we provide an automated pipeline for synthesizing choral music data from sampled instrument plugins within controllable options for instrument expressiveness. This produces an 8.2-hour-long choral music dataset from the JSB Chorales Dataset and one can easily synthesize additional data. Second, we conduct an experiment to evaluate multiple separation models on available choral music separation datasets from previous work. To the best of our knowledge, this is the first experiment to comprehensively evaluate choral music separation. Third, experiments demonstrate that the synthesized choral data is of sufficient quality to improve the model's performance on real choral music datasets. This provides additional experimental statistics and data support for the choral music separation study.
\end{abstract}
\section{Introduction}\label{sec:introduction} 

Choral music is a distinct artistic genre that includes several vocal parts (e.g.~soprano, alto, tenor, and bass) arranged into intricate patterns from strict counterpoint to polyphonic echoes and flows of lyrics. 
% Numerous works of choral music exist, including Bach's choral music collections and Barbershop compositions. 
One useful tool in the analysis and re-production of choral tracks is the ability to take mixed-down choral music and separate it back into audio tracks of isolated vocal parts: i.e. choral music separation, as a subtask of audio source separation.

Audio source separation is an audio signal processing task that involves separating one or more sound sources from a multi-source audio mixture. This task has a wide range of applications in a variety of domains, including speech separation, vocal-accompaniment separation and musical instrument separation. The latter two tasks are primary tasks in the field of music signal processing and have been adopted for practical use in the entertainment industry \cite{mss-general}. Many models, such as Open-Unmix \cite{umx}, Demucs \cite{demucs}, and Spleeter \cite{spleeter2020}, achieve great separation performance. Some models \cite{ke-sep} further extend the source separation task to a zero-shot or query-based setting. However, choral music separation has received limited attention. Unlike general musical instrument separation, which seeks to separate non-homologous sources (e.g., piano, drums, and singing voice), choral music instrument separation seeks to separate homologous or close-homologous sources (e.g., soprano, alto, tenor, bass). Additionally, the scarcity of data on choral music separation impedes further progress.
Choral music separation could be used in a wide variety of scenarios. Individuals could obtain solo tracks from choral recordings for practice, analysis, and re-production. Not only does it fill a void in a particular type of musical instrument separation, but it also provides convenience for music educators.

In this paper, we investigate the choral music separation task from the perspective of addressing the insufficiency of available datasets. We begin by introducing related works in the field of choral music separation. Second, we present the discovery of how to improve the performance of choral music separation using high-quality, synthesized music data. Then, we conduct comprehensive experiments with multiple models and datasets to evaluate the improvement of using synthesized data on choral music separation in real datasets. Finally, we discuss the extensibility of our pipeline to more choral-related separations, such as string quartet separation, as well as its future directions. The code and the dataset are publicly available\footnote{\href{https://github.com/RetroCirce/Choral\_Music\_Separation}{https://github.com/RetroCirce/Choral\_Music\_Separation}}.

\section{Related Work}
Research in choral music separation receives relatively less attention. Deep learning methods for audio source separation has already outperformed traditional methods (e.g.~Non-negative Matrix Factorization \cite{nmf}) for a long time. Separation models have been developed following two directions: frequency-domain models and time-domain separation models.
 
\subsection{Frequency-Domain and Time-Domain Models}

The traditional method of audio separation is to mask the frequency-domain representation and then inversely transform it to the time-domain signal, referred as frequency-domain separation models. Spec-U-Net \cite{unet}, based on the U-Net architecture, contains convolutional neural network (CNN) blocks for downsampling the input short-time Fourier transform (STFT) spectrogram and upsampling the bottom feature back into a separation mask. The mask is applied into the input to obtain the separate spectrogram as output. Res-U-Net \cite{resunet} replaces the original CNN blocks with residual CNN blocks to accelerate convergence speed. On the other hand, time-domain models perform the separation directly on the audio waveform. Wave-U-Net \cite{wavunet} incorporates an end-to-end U-Net structure on the input and output of waveforms. Conv-TasNet \cite{convtasnet} applies a CNN encoder-decoder structure to process waveforms into latent features and generates the mask. The masked latent features are decoded back to waveforms as separation results. Bypassing the spectrogram processing, time-domain models can save parameters and perform efficiently in low-latency systems for speech separation. Some hybrid models, such as Demucs v3 \cite{demucs}, can leverage both time-domain and frequency-domain features to achieve the best performance for musical instrument separation, while the size of the model is a little bit large.  
% However, due to the lack of  harmonic information in the spectrogram, time-domain models do not typically achieve performance parity with frequency-domain models in the field of musical instrument separation.

\begin{table}[t]
\centering
\resizebox{\columnwidth}{!}{
\begin{tabular}{lccc}
\toprule
Dataset & Minutes & Songs &Public \\ 
\midrule
Choral Singing Dataset \cite{choralesingingdataset} & 7 & 3 &$\checkmark$\\ 
Dagstuhl ChoirSet \cite{dcsdataset} & 5 & 2  &$\checkmark$\\ 
Cantoria Dataset \cite{cms-thesis}  & 20 & 14 &$\checkmark$\\ 
ESMUC Choir Dataset \cite{cms-thesis} & 31 & 26 &$\checkmark$\\ 
Bach and Barbershop Collection \cite{cms-vhs}  & 105 & 48\\ 
\bottomrule
\end{tabular}}

\caption{Existing datasets for choral music separation.}

\label{tab:q2-cms-datasets}
\end{table}

\subsection{Choral Music Separation}
For choral music separation, \cite{si-cms-man} proposed a score-informed separation model based on Wave-U-Net and performed experiments on 347 (synthesized) Bach Chorale pieces from MIDI files with \textit{MuseScore\_General} SoundFont. This model performs well on this SoundFont-Synthesis dataset but poorly on real choral music datasets. \cite{si-cms-dar} proposed a conditional Spec-U-Net to optimize the separation performance by conditioning on the fundamental frequency contour. However, as mentioned in their paper, due to the lack of choral music datasets, the evaluation was conducted on only three songs with a total duration of seven minutes. \cite{cms-vhs} proposed a harmonic overlap score to increase the model's sensitivity to different choral voices, thereby improving performance. It made use of a relatively large dataset containing 105 minutes of Bach and Barbershop Collections, but this dataset is \textit{not} publicly available due to copyright concerns, which prevents it from being open source. And indeed, 100-minute is still not enough to help achieve an audio separation model with a high generalization ability, we expect to obtain a size more than that.

In this paper, we first conduct four fundamental models: Spec-U-Net, Res-U-Net, Wave-U-Net and Conv-TasNet. Our objective is to demonstrate the efficacy of synthesized expressive data in improving separation performance on \textbf{real choral music datasets}. As a result, fundamental models enable us to consider the performance gains more directly associated with data changes and augmentations. 
% The improvement of some complex model designs (e.g. Open-Unmix \cite{umx}, Demucs \cite{demucs}, Spleeter \cite{spleeter2020}) can be considered as future work. 
Also, score-informed and conditional separation models introduce external information, such as musical notes of original songs or multi-pitch estimation results, to guide the separation's goal, while it also limits its applications. In practice, we frequently find ourselves in situations where the only available input is the audio. We continue to demand \textbf{unconditioned choral music separation}. As a result, we proceed directly to unconditioned choral music separation in this paper, without relying on any score conditions.

\begin{figure*}[t]
    \centerline{
    \includegraphics[width=\textwidth]{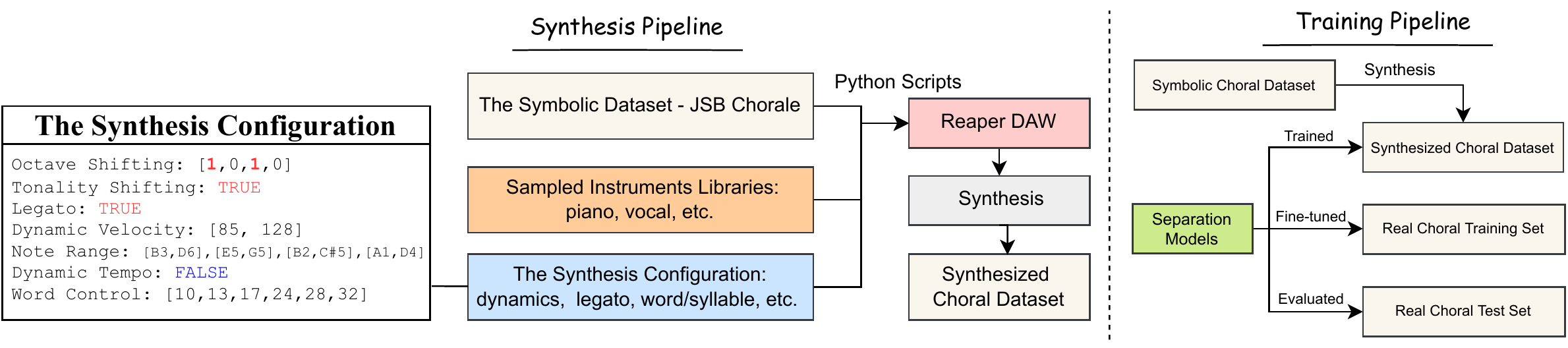}}
    \caption{The synthesis pipeline of choral music data from sample instruments and the training pipeline to utilize it.}
    \label{fig:q2-cms-pipeline}
\end{figure*}

\begin{table}[t]
\centering
\resizebox{\columnwidth}{!}{
\begin{tabular}{lcc@{~~}c@{~~}c@{~~}c}
\toprule
&&\multicolumn{4}{c}{Pitch range}\\ \cmidrule(lr){3-6}
Name        & Type & Soprano  & Alto & Tenor    & Bass        \\ \midrule
Standard MIDI   && A0--C8  & A0--C8 & A0--C8  & A0--C8 \\ 
Noire \cite{noire} & Piano      & A0--C8  & A0--C8 & A0--C8  & A0--C8 \\ 
Grandeur \cite{grandeur}  & Piano       & A0--C8  & A0--C8 & A0--C8  & A0--C8 \\ 
Voices Of Rapture \cite{vor} & Vocal     & B3--D6  & E3--G5 & B2--C\#5  & A1--D4  \\ 
Dominus Choir \cite{vodominus} & Vocal       & G3--A5  & G3--A5 & E2--G4  & E2--G4  \\ 
\bottomrule

\end{tabular}}

\caption{Sample instrument libraries we use for synthesizing choral music separation datasets.}

\label{tab:q2-sampled-instrument}
\end{table}

\section{Methodology} \label{sec:cms-methodlogy}

\subsection{Scarcity of Datasets}

Existing datasets for choral music are listed in Table \ref{tab:q2-cms-datasets}, collected from previous works and other public sources. We observe that most of these datasets have short total lengths; three of them are less than 20 minutes. The Choral Singing Dataset \cite{choralesingingdataset} and ESMUC Choir Dataset \cite{cms-thesis} have been used for choral music separation by \cite{si-cms-dar}, while the Dagstuhl ChoirSet \cite{dcsdataset} and Cantoria \cite{cms-thesis} Datasets were never used for separation tasks but instead for singing performance analysis. The Bach and Barbershop Collection \cite{cms-vhs} is relatively longer, but is not publicly available. As a result, when a model is trained and tested on such a small amount of data, its generalization and separation capabilities are severely limited. \cite{si-cms-man} directly synthesized 347 choral pieces of Bach's from MIDI files with \textit{MuseScore\_General} SoundFont and trained the model. However, this SoundFont-Synthesis dataset is dissimilar to true choral vocals. Moreover, it lacks lyrics and syllables. As a result, the trained model performs poorly on real datasets \cite{si-cms-man}.
 
In the next section, we first introduce the pipeline of synthesizing audio datasets for choral music separation from sampled instrument libraries. Then, we train various models on our datasets and compare them to determine the best model. Finally, we transfer the best model weights to the real-world datasets shown in Table \ref{tab:q2-cms-datasets}, fine-tune the model and determine whether it truly improves performance when compared to the previous settings.

\subsection{Data Synthesis Pipeline}

Figure \ref{fig:q2-cms-pipeline} shows our pipeline of choral music dataset creation and training methods. Generally, we need three collection steps:
\begin{enumerate}[leftmargin=*]
    \item The symbolic choral music dataset (MIDI, MusicXML)
    \item The sampled instrument libraries (standalone VST plugin, or Kontakt sample libraries)
    \item The synthesis configuration (syllables or lyrics choices, legato, velocity, and tempo)
\end{enumerate}
Then, our provided code can completely automate the data synthesis process. It is built on top of the python-support and free digital audio workstation (DAW) -- Reaper\footnote{https://www.reaper.fm/}. With the above three steps, one could complete any choral music data synthesis process on supported system platforms.

\subsection{Data and Instrument Collection}
For Step 1, following \cite{si-cms-man}, we use the JSB Chorales Dataset \cite{jsbchorale}, which contains 347 pieces of choral music in MusicXML format. The total duration is 248 minutes at a tempo of 90 bpm (a.k.a. beat per minute). The data is first transformed into MIDI files, which serves as the symbolic dataset source for the creation of choral music audio. For Step 2, a sampled instrument\footnote{A detailed introduction can be found at https://tinyurl.com/2p8trn2u} is a sound source plugin applied in a DAW. Unlike a SoundFont, it contains samples of real instruments recorded in a professional acoustic environment. The human singing voice is also considered as an instrument type. And many vocal sampled instruments support a variety of lyrical or syllabic sets (e.g., vowels, Latin words, etc.).  We first choose two types of instruments for our purposes: piano and vocal (soprano, alto, tenor, and bass). Then, we choose two sampled instruments for each type, as shown in Table \ref{tab:q2-sampled-instrument}.
% The four columns on the right indicate the pitch range of each instrument in the library.
The reason to choose the piano instrument is to evaluate the separation performance of piano as a common and same-source instrument.
% , even though in the real-world we never provide separated tracks of each ``voice'' in piano pieces. 
In this case, the model needs to consider most on the pitch difference between each voice part.
When it comes to the vocal dataset, the model can distinguish the timbres slightly between soprano, alto, tenor, and bass voices, but it is more difficult to model the acoustic features of these four voices than piano. This allows us to determine whether the model can improve performance by exploiting the timbre difference between vocal datasets, or if it fails to model these timbres and perform a bad separation result.

Due to the fact that we have two sampled libraries for each type of instrument, each dataset contains 248 $\times$ 2 = 496 minutes (8.2 hours) of synthesized choral music data. All sampled instrument libraries that we use have a paid license.

\subsection{Synthesis Configuration for Expressiveness}  \label{sec:expressive-config}
For Step 3, we adopt two methods to further improve the scalability and quality of synthesized data: the basic data augmentation, and expressiveness incorporation. The left of Figure \ref{fig:q2-cms-pipeline} shows a specification.
%on it.

We perform two operations to augment the data. First, we notice that the pitch ranges of sampled instrument libraries do not always correspond exactly to the pitch ranges of tracks in JSB Chorales Dataset. For instance, some bass melodies in JSB appear to be lower than the lowest note in sampled instrument libraries. Instead of directly discarding these tracks, we implement ``octave shifting'' by shifting out-of-range notes up or down some octaves until they fall within the range. While it produces non-realistic jumps between some melodies, it saves the whole track to preserve more realistic data. Second, we apply ``tonality shifting''
% \footnote{While  ``tonality shifting'' is lossless in the symbolic domain prior to synthesis, it is lossy and thus a difficult augmentation to achieve on existing real audio datasets.}
to each track. The tonality was shifted upward and downward by three semitones before synthesis. Therefore, the effective length of training data will be further augmented several times. 

For expressiveness incorporation, we provide several options, which are supported by sampled instrument libraries, to synthesize audio:
\begin{itemize}[leftmargin=*]
    \item Legato: for vocal, this includes whether or not to change breath or sing continuously. In vocal instrument libraries, legato is controlled by detecting the presence of an overlap between adjacent notes. To support the legato configuration, we begin by segmenting the track into \textit{musical phrases} using the breath break information in MusicXML (if provided) or the note intervals (if specified). Then, in each phrase, we add overlap to adjacent notes (to activate the legato) if their pitch distance is less than 7 semitones (i.e., a perfect fifth).
    \item Velocity: for each phrase, we provide three types of volume/velocity change curves: \textit{crescendo}, \textit{diminuendo}, and \textit{cresc.$\rightarrow$dim.}. The configuration establishes the maximum and minimum velocity ranges. 
    \item Word Control: for vocal, we support the word control of sampled instrument libraries by assigning random combinations of words or syllables to each phrase. Note that real-world performance may not contain random word changes, but for  model training, this still increases the data richness on each training batch.
\end{itemize}
The configuration also supports the reverberation as a designed feature, but currently it is not applied in this work.

\section{Experiments}

In this section, we conduct an experiment on evaluating different separation models on our synthesized datasets. The purpose of this experiment is to identify the best model on synthesized datasets and then transfer it to real choral music datasets.

\subsection{Datasets, Models and Hyperparameters} \label{sec:dataset-model-hyper}
As introduced above, we use two datasets (piano and vocal) to train four models (Spec-U-Net, Res-U-Net, Conv-TasNet, and Wave-U-Net). Each track is in 22,050 Hz sample rate. Each dataset contains 496 minutes data. We use 277 tracks for training, 35 tracks for validation and another 35 tracks for testing. Since we have training combinations among four models, two datasets, and four choral voices, to save training time, we first train models on the dataset \textit{without} the expressiveness incorporation in section \ref{sec:expressive-config}, named as \textbf{Standard-Piano} and \textbf{Standard-Vocal} datasets. After finding the best model, we will train it on the expressive datasets in section \ref{sec:finetune-exp}.

For model hyper-parameters, In Spec-U-Net \cite{unet}, we use a window size of 2048, FFT size of 2048, and hop size of 441. We apply 7 CNN blocks to downsample the input spectrogram, and another 7 CNN blocks to upsample it into the separation mask. In Res-U-Net, we apply the implementation from \cite{resunet}, with 10 residual CNN blocks to downsample the input spectrogram, then another 6 residual CNN blocks to upsample. In Wave-U-Net, we follow the settings of \cite{wavunet} to adopt 6 downsampling CNN layers and 6 upsampling CNN layers for separation. The filter channels are set from 32 to 1024 in order for each layer. The kernel size is 15 for the first layer and 5 for remaining layers. In Conv-TasNet, we follow the setting of \cite{convtasnet} to set hyper-parameters as $N=512$, $L=20$, $B=128$, $H=512$, $P=3$, $R=3$, $X=8$. Spec-U-Net and Res-U-Net use their default mean absolute error (MAE) loss function; Conv-TasNet uses the default scale-invariant source-to-noise ratio (SI-SDR) loss; and Wave-U-Net with mean squared error (MSE) loss.

For training hyperparameters, the batch size is 8, the learning rate is 1e-3, and each training sample is a 2-sec audio segment randomly chosen from one music track in the training set. The number of steps for each epoch is 700. We apply the  Adam optimizer \cite{adam} with $\beta_1=0.9$, $\beta_2=0.999$, $\epsilon=1e-8$, and a learning rate scheduler where the learning rate is reduced with a multiplier $f=0.65$ if the validation performance does not improve across 3 consecutive epochs. We implemented all methods in Pytorch using NVIDIA RTX 2080Ti GPUs. All models converged within 300 epochs with early stop using a 10-epoch patience. 

For evaluation, source-to-distortion ratio (SDR) is one of the most widely used metrics for evaluating a source separation system's output, which measures a ratio between the original source track and the noise, interference, added artifacts in the separation track. It is considered to be an overall measure of how good a separation result sounds. We follow the music separation campaign SiSEC 2018 \cite{sisec2018} to use the median SDR to evaluate separation performance. The median SDR is obtained by first computing segment-level SDR of each 2-sec segment in each track, then taking the median over them as track-level SDRs, finally taking the median over the track-level SDRs as the final SDR. The computing library is \textit{mus\_eval} \cite{sisec2018}.
% For bibtex users:

\begin{table}[t]
\centering
\resizebox{\columnwidth}{!}{
\begin{tabular}{ccc@{~~}c@{~~}c@{~~}c@{~~}c}
\toprule
Standard &  \multirow{2}{*}{Model} & \multicolumn{5}{c}{Median Source-Distortion Ratio (dB)} \\ \cline{3-7}
Dataset & & Soprano & Alto & Tenor & Bass & Avg.   \\ \midrule
Piano & Spec-U-Net\cite{unet} & \textbf{9.78} & \textbf{9.46} & \textbf{10.35} & 10.60 &  \textbf{10.05}  \\
Piano & Res-U-Net\cite{resunet} & 8.53 & 9.01 & 9.97 & \textbf{12.23} & 9.94 \\
Piano & Wave-U-Net\cite{wavunet} & 6.95 & 5.36 & 7.21 & 9.82 & 7.34    \\
Piano & Conv-TasNet\cite{convtasnet} &  7.04 & 6.98 & 7.29 & 7.82 & 7.28 \\ 
\midrule
Vocal & Spec-U-Net\cite{unet} & \textbf{10.45} & 10.19 & \textbf{12.25} & 9.53 &  \textbf{10.61} \\ 
Vocal & Res-U-Net\cite{resunet} & 9.35 & \textbf{10.87} & 10.20 & \textbf{10.77} & 10.30 \\ 
Vocal & Wave-U-Net\cite{wavunet} & 2.65 & 3.08 & 3.06 & 3.90 & 3.17   \\
Vocal & Conv-TasNet\cite{convtasnet} & 6.60 & 6.12 & 6.41 & 6.58 & 6.43 \\ 
\bottomrule
\end{tabular}}
\caption{The separation performance of four models on the test sets of Standard-Piano and Standard-Vocal datasets.}
\vspace{-0.5cm}
\label{tab:q2-csm-standard-performance}
\end{table}

\begin{table*}[t]
\centering
\resizebox{0.85 \textwidth}{!}{
\begin{tabular}{ccccc}
\toprule
\multirow{2}{*}{Fine-Tuning Evaluation Set} & \multirow{2}{*}{Pretraining Set} & \multicolumn{3}{c}{Avg. Median SDR (Fine-Tuning Ratio)} \\ \cline{3-5}
 &  & \multicolumn{1}{c}{ratio=10\%} & \multicolumn{1}{c}{ratio=40\%} & \multicolumn{1}{c}{ratio=70\%} \\ \midrule 
\multirow{4}{*}{Cantoria Dataset \cite{cms-thesis}}  & \multicolumn{1}{|c}{None}  & \multicolumn{1}{|c|}{1.42} & \multicolumn{1}{c|}{3.91} & 4.13 \\ 
  &  \multicolumn{1}{|c}{SoundFont-Synthesis \cite{si-cms-man}} & \multicolumn{1}{|c|}{2.39} & \multicolumn{1}{c|}{3.90} & 4.03 \\ 
 & \multicolumn{1}{|c}{Standard-Vocal (ours)}  & \multicolumn{1}{|c|}{3.03} & \multicolumn{1}{c|}{4.59} & 5.08 \\ 
 & \multicolumn{1}{|c}{Expressive-Vocal (ours)}  & \multicolumn{1}{|c|}{3.73} & \multicolumn{1}{c|}{5.48} & 5.71 \\ \midrule 
\multirow{4}{*}{Choral Singing Dataset (CSD) \cite{choralesingingdataset}} & \multicolumn{1}{|c}{None} & \multicolumn{1}{|c|}{1.98} & \multicolumn{1}{c|}{2.78} & 5.26 \\ 
 & \multicolumn{1}{|c}{SoundFont-Synthesis \cite{si-cms-man}} & \multicolumn{1}{|c|}{2.12} & \multicolumn{1}{c|}{3.38} & 6.20 \\ 
 & \multicolumn{1}{|c}{Standard-Vocal (ours)} & \multicolumn{1}{|c|}{3.43} & \multicolumn{1}{c|}{4.23} & 6.91 \\ 
 & \multicolumn{1}{|c}{Expressive-Vocal (ours)}  & \multicolumn{1}{|c|}{4.19} & \multicolumn{1}{c|}{4.78} & 7.50 \\ \midrule
\multirow{4}{*}{Bach \& Barbershop Collection (BBC) \cite{cms-vhs}} & \multicolumn{1}{|c}{None}  & \multicolumn{1}{|c|}{4.18} & \multicolumn{1}{c|}{6.08} & 6.94 \\ 
 & \multicolumn{1}{|c}{SoundFont-Synthesis \cite{si-cms-man}} & \multicolumn{1}{|c|}{4.19} & \multicolumn{1}{c|}{6.17} & 6.93 \\
 & \multicolumn{1}{|c}{Standard-Vocal (ours)}  & \multicolumn{1}{|c|}{4.98} & \multicolumn{1}{c|}{6.71} & 7.27 \\ 
 & \multicolumn{1}{|c}{Expressive-Vocal (ours)}  & \multicolumn{1}{|c|}{5.58} & \multicolumn{1}{c|}{7.17} & 7.64 \\ \bottomrule
\end{tabular}
}
\caption{The fine-tuning performance of three real datasets by our best model -- Spec-U-Net.}
\vspace{-0.5cm}
\label{tab:q2-csm-few-shot-avg}
\end{table*}

\subsection{Separation Performance}

Table \ref{tab:q2-csm-standard-performance} shows all four parts median SDR performance on two standard datasets by four models. We can see that the frequency-domain models Spec-U-Net and Res-U-Net get similar results that are better than those of the time-domain models Conv-TasNet and Wave-U-Net. Spec-U-Net achieves the best average SDRs over four parts on two datasets as 10.05 and 10.61. The Res-U-Net achieves very close performance. When analyzing the results, frequency-domain models can take advantage of spectrograms in choral music to obtain better separation results. The performance on soprano, alto and tenor in vocal is better than that in the piano dataset, suggesting that the timbre difference can also help further discriminate different source tracks. Time-domain models can model the piano acoustic features well to achieve a good performance, but find it hard to model the vocal features solely on the waveform and face the drops in the vocal dataset.

\begin{figure*}[t]
    \centering
    \begin{subfigure}[b]{\textwidth}
        \centering
        \includegraphics[width=0.88\textwidth]{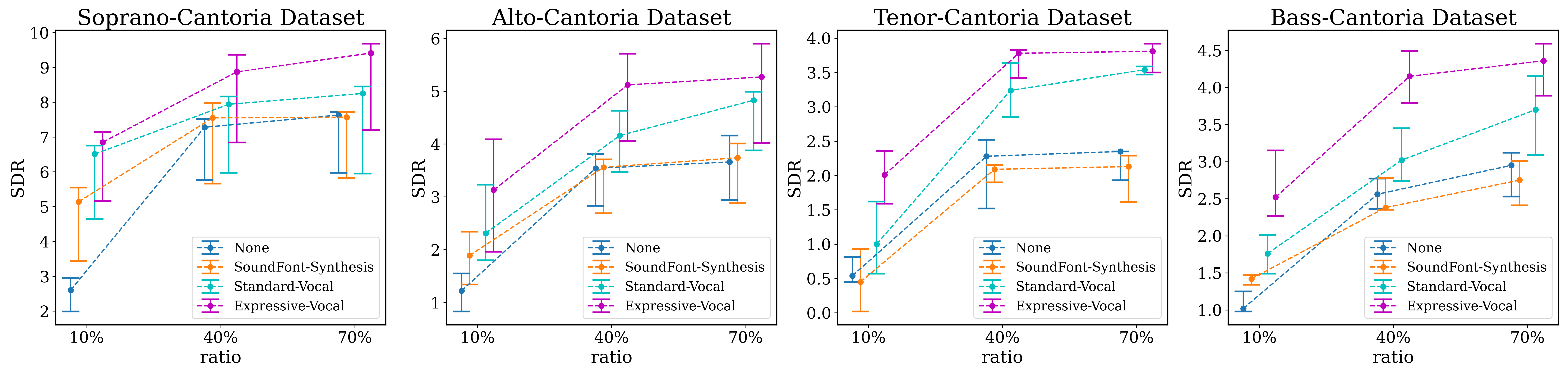}
        \caption{The Median SDR performance on the Cantoria Dataset.}
        % \label{fig:q2-few-shot-chart-ctd}
    \end{subfigure}

    \begin{subfigure}[b]{\textwidth}
        \centering
        \includegraphics[width=0.88\textwidth]{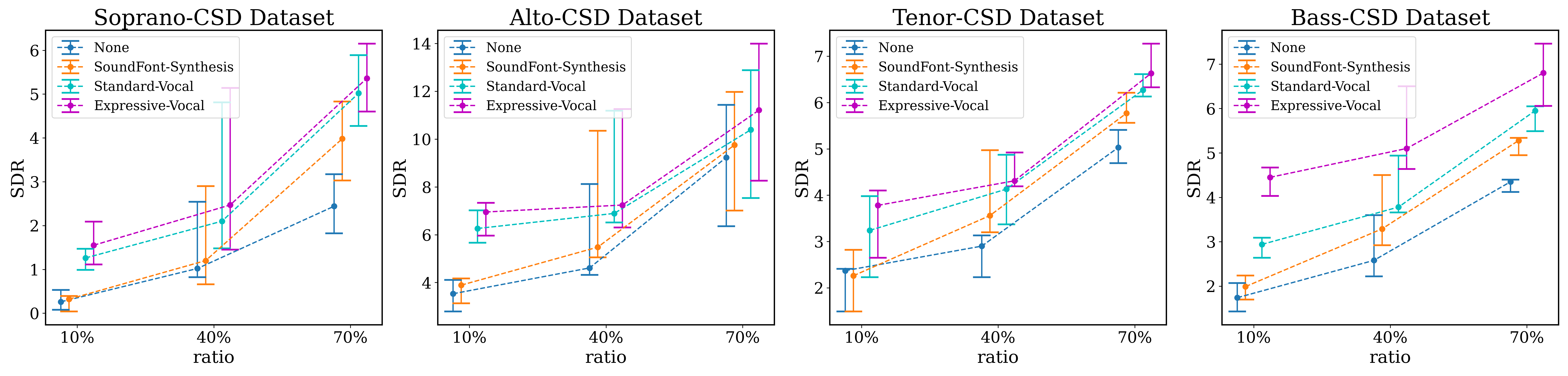}
        \caption{The Median SDR performance on the Choral Singing Dataset (CSD).}
        
    \end{subfigure}

    \begin{subfigure}[b]{\textwidth}
        \centering
        \includegraphics[width=0.88\textwidth]{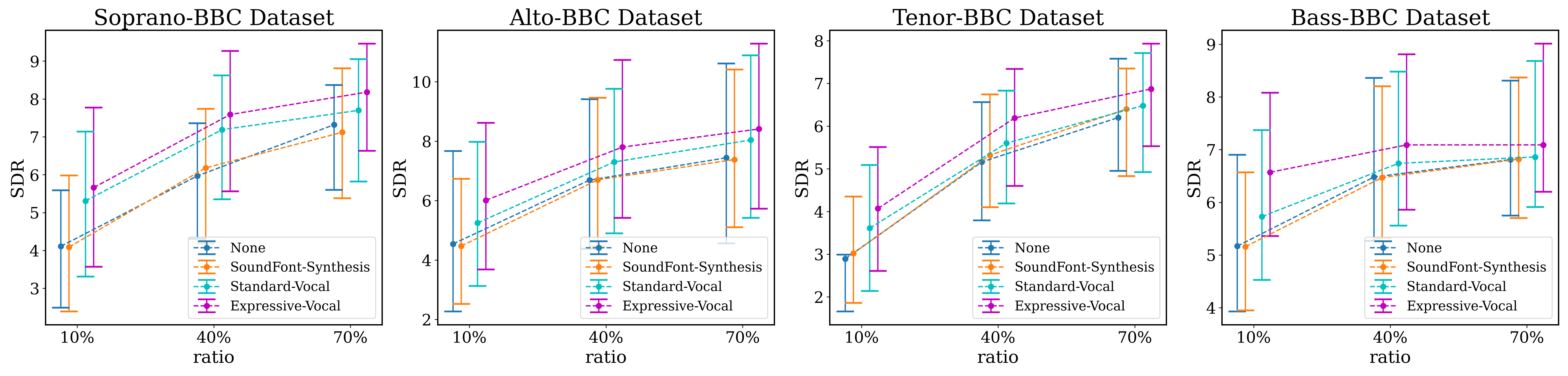}
        \caption{The Median SDR performance on the Bach \& Barbershop Collection Dataset (BBC).}
        % \label{fig:q2-few-shot-chart-bbc}
    \end{subfigure}
    \caption{The Median SDR performance, with a 25th--75th percentile range, of soprano, alto, tenor and bass on three datasets by Spec-UNet with different pretrained models and different ratios of training-test sets.}
   \label{fig:q2-few-shot-chart}
   \vspace{-0.4cm}
\end{figure*}

\subsection{Fine-Tuning Evaluation on Real Datasets} \label{sec:finetune-exp}
After comparing models in standard datasets, we chose the best model, Spec-U-Net, to conduct the next experiments. We trained Spec-U-Net on the \textbf{Expressive-Vocal} dataset, as we synthesized the data \textit{with} the expressiveness incorporation. Then, as shown in the right of Figure \ref{fig:q2-cms-pipeline}, we saved the best model checkpoints, and conducted a fine-tuning experiment to verify whether our data is useful for transfer learning on real choral music datasets. Table \ref{tab:q2-csm-few-shot-avg} and Figure \ref{fig:q2-few-shot-chart} illustrate the median SDR performance of three real choral music datasets under different fine-tuning ratios with different pretrained models. 

For datasets, we chose the Cantoria Dataset, Choral Singing Dataset (CSD), and Bach \& Barbershop Collection\footnote{We appreciate the help from authors in \cite{cms-vhs} to offer the dataset.} (BBC). The reason for these choices is that Cantoria contains the best recording quality, CSD is most frequently used in previous works, and BBC contains the longest length. The meta information of each dataset has been described in Table \ref{tab:q2-cms-datasets}.

We considered three ratios for fine-tuning: (1) 10\% for training, 90\% for evaluation; (2) 40\% for training, 60\% for evaluation; and (3) 70\% for training, 30\% for evaluation. The fine-tuning experiments demonstrate if our synthesized datasets can improve the separation performance in real datasets under different settings (e.g., few-shot as 10\% and fairly enough as 70\%). The intermediate ratio 40\% is conducted to further investigate the tendency of the improvement brought by our datasets.   
% Three datasets contain different lengths of data, as a general statistics, 10\% data is ranged from 0.7--10 minutes, 40\% data is from 3--42 minutes, and 70\% data from 5--73 minutes.

There are four dataset choices on which to pretrain the models: (1) None: without any pretraining; (2) SoundFont-Synthesis: the synthesis dataset in \cite{si-cms-man} by the \textit{Musescore\_General} SoundFont as a baseline; (3) the Standard-Vocal dataset; and (4) the Expressive-Vocal dataset. Since the SoundFont-Synthesis dataset only contains 248 minutes, instead of using two sampled libraries (496 minutes), we only provide the data synthesized from one library -- Voices Of Rapture \cite{vor} in Standard-Vocal and Expressive-Vocal for the pretraining. Data augmentations of ``octave shifting" and ``tonality shifting" are applied in all three datasets, except (4) incorporates more expressiveness settings. The fine-tuning learning rate is 1e-4, with the scheduler in section \ref{sec:dataset-model-hyper}.

Table \ref{tab:q2-csm-few-shot-avg} shows the average median SDR performance of Spec-U-Net over four voice parts under different fine-tuning ratios and different pretraining settings. We can see that under all three training-test ratios, the performance of the model pretrained on Standard-Vocal and Expressive-Vocal is better than that on SoundFont-Synthesis and none-dataset, where the performance of Expressive-Vocal achieves the best. When the training-test ratio is small as 10\%, the performance of SoundFont-Synthesis and non-dataset has the largeest difference, showing that the model learns some priors from SoundFont-Synthesis and converges to a better optimum. However, when the ratio increases to 40\% and 70\%, their performance is close to each other and does not vary much, especially on Cantoria and BBC. Thus, pretraining on SoundFont-Synthesis dataset provides a very useful initialization -- but gains diminish (or even no gain) as the initializer is dominated by larger and larger quantities of real training data.

However, when the model is pretrained on Standard-Vocal, it has a strong generalization to real choral music datasets under all fine-tuning ratios, because acoustic features of synthesis tracks share large similarity to the real datasets. This performance is further boosted by Expressive-Vocal as we introduce expressiveness during synthesis, such as lyrics and velocity dynamics. Even under the 70\% fine-tuning ratio, as the model has received many real data, Standard-Vocal and Expressive-Vocal pretrained model can still get improvements. In conclusion, our synthesized datasets provide not only additional data volume, but also high-quality and close-to-real choral music samples for boosting the separation performance. 

To further verify our analysis, we visualized the trends of median SDRs (blue, orange, cyan, and magenta colors), with a 25th-75th percentile range, for each voice part of three real datasets in Figure \ref{fig:q2-few-shot-chart}. We can see the performance of our synthesized datasets (magenta \& cyan lines) marks a clear performance increase and a large gap to that of SoundFont-Synthesis and non-dataset (orange \& blue lines). However, the trends of SoundFont-Synthesis and non-dataset are close to each other, and even overlap in BBC. When analyzing the percentile range of each model, on Cantoria and CSD, our Standard-Vocal and Expressive-Vocal pretrained models reveal a clear difference of percentile ranges to the left two models, demonstrating that our models get a large improvement. However, the percentile ranges of the SoundFont-Synthesis and non-dataset pretrained models have a large overlap, demonstrating no difference even though their median SDRs differ a little. On the BBC dataset, our models yield improvements on both the 25th and 75th percentile values but are not as pronounced as those observed on Cantoria and CSD. The potential reason is because the BBC dataset contains a relatively large data size (105 minutes), which makes the model already achieve a good convergence and hard to get more significant improvements without model designs. These trends further demonstrates that our synthesized dataset plays a role in making up the data scarcity and improving generalization ability.

\section{Extensibility and Limitations}
Our provided synthesis pipeline from symbolic datasets to real audio datasets not only benefits choral music separation tasks, but also other choral-related separation tasks. For example, string quartet separation, to separate two violins, viola, and cello parts from a mixed audio, can also be trained with synthesized data of our pipeline. The details of the string quartet separation experiment can be accessed in the code repository. Similarly, our best pretrained model shows a 100\%/30\% performance increase to the SoundFont-Synthesis and non-dataset pretrained models.  This further shows a potential application of our synthesis pipeline to improve other choral-related separation tasks.

% As an additional experiment, we trained Spec-U-Net models to separate string quartets using Standard-String and Expressive-String datasets, synthesized from JSB Chorales Dataset by a string sampled instrument library -- Intimate Strings \cite{iistring}. 

% The word changes, when applied into string instruments, become the playing mode changes, namely \textit{sustain}, \textit{tremolo}, and \textit{pizzicato}. We conducted the same fine-tuning experiments with 10\% and 40\% fine-tuning ratios on a 8-minute real string quartet subset in URMP dataset \cite{urmpdataset}. Under the 10\%/40\% fine-tuning ratios, our best pretrained model shows a 100\%/30\% performance increase to the SoundFont-Synthesis and non-dataset pretrained models.

There are also some limitations and future improvements to our work. First of all, our implementations of expressiveness are still based on random template modes. Deep learning methods can improve this expressiveness modeling \cite{midi-ddsp}. Second, the design of the choral separation model is needed to learn more priors from weak synthesized data that can be transferred to real data, then it will complement our proposed pipeline better. These limitations are planned for exploration in our future work.

\vspace{-0.2cm}
\section{Conclusion}
In this paper, we proposed an automated pipeline for synthesizing choral music data from sampled instrument plugins, and created an 8.2-hour choral music dataset to improve separation performance on real choral music datasets. We comprehensively evaluated multiple separation models to demonstrate that synthesized choral data is of sufficient quality to improve model's performance on real datasets. This provides additional experimental statistics and data support for choral music separation study. In the future, we will focus on the design of timbre-pitch disentanglement model \cite{ke-sketchnet} for achieving better separation performance. The application of choral music separation results into other music-related tasks, such as music recommendation \cite{ke-recom}, is also planned as the future work.

\bibliography{ISMIRtemplate}

% For non bibtex users:
%\begin{thebibliography}{citations}
% \bibitem{Author:17}
% E.~Author and B.~Authour, ``The title of the conference paper,'' in {\em Proc.
% of the Int. Society for Music Information Retrieval Conf.}, (Suzhou, China),
% pp.~111--117, 2017.
%
% \bibitem{Someone:10}
% A.~Someone, B.~Someone, and C.~Someone, ``The title of the journal paper,''
%  {\em Journal of New Music Research}, vol.~A, pp.~111--222, September 2010.
%
% \bibitem{Person:20}
% O.~Person, {\em Title of the Book}.
% \newblock Montr\'{e}al, Canada: McGill-Queen's University Press, 2021.
%
% \bibitem{Person:09}
% F.~Person and S.~Person, ``Title of a chapter this book,'' in {\em A Book
% Containing Delightful Chapters} (A.~G. Editor, ed.), pp.~58--102, Tokyo,
% Japan: The Publisher, 2009.
%
%
%\end{thebibliography}

\end{document}